\documentclass[manuscript, nonacm]{acmart}

\usepackage{array}

\AtBeginDocument{%
  }

\setcopyright{acmlicensed}
\copyrightyear{2027}
\acmYear{2027}
\acmDOI{XXXXXXX.XXXXXXX}
\acmConference[Conference acronym 'XX]{Make sure to enter the correct
  conference title from your rights confirmation email}{June 03--05,
  2018}{Woodstock, NY}
\acmISBN{978-1-4503-XXXX-X/2018/06}

\begin{document}

\title[Working with Agentic `Teammates']{Working with Agentic `Teammates': When a New Organizational Actor Collides with the Human Ecosystem of Work}


\settopmatter{authorsperrow=4}
\author{Rida Qadri}
\authornote{Both authors contributed equally to this research.}
\author{Remi Denton}
\authornotemark[1]
\authornote{Corresponding author: dentone@google.com}
\affiliation{%
  \institution{Google Research}
  \country{USA}
}

\author{Michael Madaio}
\affiliation{%
  \institution{Google Research}
  \country{USA}}

\author{Mahima Pushkarna}
\affiliation{%
  \institution{Google DeepMind}
  \country{USA}
}

\author{Leslie Lai}
\affiliation{%
  \institution{Google Research}
  \country{USA}
}

\author{Sherry Moore}
\affiliation{%
  \institution{Google DeepMind}
  \country{USA}
}

\author{Michelle Chen Huebscher}
\affiliation{%
  \institution{Google DeepMind}
  \country{Switzerland}
}

\author{Andrew Butcher}
\affiliation{%
  \institution{Google DeepMind}
  \country{USA}
}

\author{Ritom Sen}
\affiliation{%
  \institution{Google DeepMind}
  \country{USA}
}

\author{Hsiao-Yu Tung}
\affiliation{%
  \institution{Google DeepMind}
  \country{USA}
}

\author{Shaan Mathur}
\affiliation{%
  \institution{Google DeepMind}
  \country{USA}
}

\author{Yimeng Liu}
\affiliation{%
  \institution{Google DeepMind}
  \country{USA}
}

\author{Edward Grefenstette}
\affiliation{%
  \institution{Google DeepMind}
  \country{United Kingdom}
}

\author{Noah Fiedel}
\affiliation{%
  \institution{Google DeepMind}
  \country{USA}
}

\author{Shibl Mourad}
\affiliation{%
  \institution{Google DeepMind}
  \country{Canada}
}

\author{Michael Terry}
\affiliation{%
  \institution{Google DeepMind}
  \country{USA}
}
\renewcommand{\shortauthors}{Qadri and Denton et al.}
\newcommand{\agent}{Team Agent}
\newcommand{\pquote}[1]{``\textit{#1}''}

\begin{abstract}
Enterprise AI is transitioning from single-user, reactive tools toward proactive, multi-user ‘teammates,’  but our empirical understanding of this transition is limited. In this paper, we  present an in-situ qualitative study of a persistent, proactive AI agent `teammate' deployed across multiple teams in a large technology company.  Our findings reveal the boundaries of the human-agent workplace are actively in flux, triggering breakdowns and negotiations across: 1) tacit rules of collaborative human workflows, 2) the relational boundaries of this new non-human actor, and 3) the redistribution of trust and human agency. We use these early micro-negotiations as signals to chart a new research, design, and organizational agenda  that intentionally preserves human agency in a workplace shared with non-human organizational actors.




\end{abstract}

\received{20 February 2007}
\received[revised]{12 March 2009}
\received[accepted]{5 June 2009}

\maketitle

\section{Introduction}

The paradigm of artificial intelligence is undergoing a fundamental shift from single-user, reactive tools awaiting instruction, to multi-user, autonomous systems. In the organizational enterprise context, these systems are increasingly positioned as proactive `teammates' capable of autonomous execution  (e.g., Claude Tag \cite{claudetag}, Grok Bot \cite{grokbot}, Slackbot \cite{slackbot}). These `teammates' can be   embedded directly into shared collaborative surfaces such as team chats and documents, and can ambiently monitor team context, decompose high-level objectives, interface directly with existing digital infrastructure (such as databases, code repositories, and collaborative software), and execute complex, multi-step workflows in the background.    
As this new class of highly autonomous agents is being deployed into an ecosystem of work historically built by and for humans,   critical questions emerge about the future of human-AI interaction in the workplace: How should people conceptualize and relate to these agents? What happens when this new class of autonomous systems are deployed in the wild? What protocols should govern their behavior? How should their integration be designed? 

While nascent HCI literature has begun studying this emerging paradigm,  our understanding of interaction with AI `teammates' has largely been informed by simulated laboratory environments and artificially bounded, short-term tasks  \cite{Zhang2023communication, nixon2026socialcostaiteammate, Schelble2026, Hu2026BossOrBot, Chung2026CodingAgent}. 
While these studies offer valuable foundations, a gap remains in the field's understanding of how agentic AI integrate into  the messy, tacit realities of collaborative work.  
Real-world, \textit{in situ} studies are critical because decades of CSCW and organizational sociology demonstrate that technologies are never adopted in a vacuum; they inevitably disrupt and reconfigure the entrenched social practices, relational norms, and unwritten conventions that underpin collaboration (and are mutually reconfigured by them in turn)  \cite{daft1984toward, suchman1987plans, schoeneborn2011organization}. Thus to understand this new workplace technology, it is useful to take a relational approach that recognizes that organizations, human practices, and collaborative technologies continuously co-evolve in practice \cite{Orlikowski2010} and study these agents within the ecosystem they are entering. \looseness=-1

To address this gap, we study the deployment of a highly proactive, persistent, autonomous agent (henceforth \agent{}) within a large technology company. This is  a  real-world instantiation of the emerging agentic `teammate' paradigm being deployed commercially.  Designed to streamline communication and coordination within collaborative work environments (e.g., by helping teams manage information flows or file bugs), \agent{} was deployed across over 20 teams and integrated into their respective collaborative spaces.  Notably, our study does not evaluate \agent{} as a finished product, nor do we obtain user feedback on specific features. Rather, rooted in the tradition of technology probes \cite{Hutchinson_probe, Wallace_probe}, we use \agent{} to explore how the design paradigm of autonomous team-embedded agents interacts with the realities of human teamwork, and what this might tell us about the future of human-agent collaboration as well as human-\textit{human} collaboration. 
Through interviews with \agent{} users, we found the agent's introduction collided with three aspects of the ecosystem of collaborative work: 
\begin{enumerate}
    \item  \textbf{Collaborative workflows and work cultures:} Despite its broad technical capabilities, \agent{}  struggled to decode the implicit, contextual practices that underpin teamwork, triggering disruptions to  work. 
  \item \textbf{Relational norms and categories:} With no pre-existing mental model for this new entity, users continuously engaged in fragmented, individual negotiations over what \agent{} was, what constituted appropriate relational boundaries, and its social positioning in the organization. 
  \item \textbf{Agency and trust}: Acceptance of \agent's proactivity and autonomy was mediated by how users  perceived it might impact their own agency, and a sense that the agent's agency should be a progressively earned privilege, rather than a  set of capabilities the agent can fully exercise from the start.
\end{enumerate}
    
Our study offers a rich snapshot of how human teams actively configure and react to a proactive AI agent within a live organizational environment. The spectrum of reactions we observed indicates a socio-technical environment in flux, where the boundaries and norms of human-agent collaboration have yet to stabilize and are actively being written. In this critical moment of transition, our study offers early empirical signals of the breakdowns that may occur and need to be paid attention to when a non-human system is granted the privileges and agency of a human teammate, despite it inherently lacking the social obligations, relational capital, and tacit knowledge required to navigate that role appropriately. Rooted in the tradition of socio-materiality   \cite{Orlikowski2010}, we argue for the value of conceptualizing agents as novel non-human organizational actors that actively reconfigure the social and material realities of work. Treating them as such demands a research and design agenda focused on the creation of new organizational structures, interaction paradigms,  and governance mechanisms that respond to this new reality where a non-human agentic actor is working alongside us.

\section{Related Work}

\subsection {The Paradigm of Human AI Teaming}
Over the last few years, the design of LLM-based artificial intelligence for the workplace has been dominated by a single-user paradigm. Operating primarily as passive, direct-command tools, systems like conversational chatbots and coding co-pilots were designed to augment individual productivity and compress skill gaps \cite{noy2023, Brynjolfsson2025} with literature examining its impacts on workers and future of work \cite{ehsan2026future}. However, an increasing body of research and commercial deployment is pivoting toward collaborative, multi-user systems where AI operates alongside groups of human workers. In this emerging paradigm, the AI is integrated directly into the team structure as a `teammate' \cite{hkuds_agentspace_2026, YanZhuZhuEnrWan26, QuaAlbWuDinChi26, SheWanCheSheHua26, ZhuThaTsaWexQia26} and is used by multiple users at the same time \cite{SwaZhoParJeoZim25, FuAnaChe24, ZhaWuRenTurMen26, KobDevParBanFer24, LeeChoMut25}. 

Defining what exactly constitutes an AI `teammate' however, remains theoretically contentious. 
For example, \citet{hughes2026types} evaluate 53 experimental studies on human-AI teaming, revealing that researchers apply the `teammate' label to very different configurations, with nearly half of the AI teammate literature examining what they call an `assistant' paradigm characterized by human-AI dyads with sequential task execution and a hierarchical command-and-control framework where the user always initiates the action. 
This dyadic, hierarchical interaction does not fit emerging definitions of AI Teammates nor the industry standard of what is being deployed in real world settings.  Distinguishing human-AI teaming from ``human-AI interaction or collaboration,'' \citet{wang2025adaptivehumanagentteamingreview} argue that the human and AI system have to work interdependently, with the AI system also actively supporting ``shared goals'' of teamwork to ``foster social and interpersonal dynamics.''  \citet{hughes2026types} also ask if the concept of a teammate muddies the water by conceptualizing human-AI teams within a framework built primarily for human-only teams. 
We use the term teammate then in this paper without valorizing or implying personhood, but adopt the term to recognize this growing paradigm of AI development. 

In contrast to the human-AI teaming structures being studied in empirical literature, \agent{} is much closer to the industry standards of what are being launched: multi-party system with operational agency to propose, structure, and drive work with a dedicated service account and 'identity' integrated into workspaces with its combination of multi-surface persistence, hybrid group and individual communication, and active social presence. 

Emerging HCI literature has recently begun to investigate the possible dynamics and impacts of human-AI teams. Methodologically, these prior works overwhelmingly examine AI `teammates' instantiated through single-user LLM-based chatbots \cite{Flathmann2025trust, nixon2026socialcostaiteammate}, a Wizard of Oz approach \cite{WOZ} (where the AI `teammate' is simulated by a human) \cite{Leong2024dittos, Duan2026trust, Schelble2026}, or video game bots \cite{Zhang2023communication, Zhang2024VerbalVsVisual, FlathmannTeammateDivide}.
Experimental setups predominantly leverage session-scoped, controlled environments such as gaming platforms \cite{Zhang2023communication, Zhang2024VerbalVsVisual, FlathmannTeammateDivide}, flight simulators \cite{Duan2026trust, Schelble2026}, or custom text-based tasks \cite{nixon2026socialcostaiteammate}. 
The teams being studied are often groups of strangers brought together for the purpose of the study and are not long-standing, real functioning teams. Using these methods these works have explored how teams calibrate trust \cite{Zhang2023communication, Duan2026trust} and form mental models \cite{FlathmannTeammateDivide} of AI `teammates' and how AI `teammates' can impact human-human communication dynamics \cite{nixon2026socialcostaiteammate}.
At the same time, qualitative exploratory and speculative studies have explored the conceptual bounds and possibility space of AI `teammates,' such as by eliciting perspectives on possible roles and expectations for AI teammates \cite{Vishwarupe2026, Chen2026conflcit}.

The ability to deeply study impacts of AI teammates in real work contexts is currently constrained by the artificiality of many study environments and the configuration of the AI-human interaction. As \citet{hughes2026types} argue, the lifespan of an interaction between an AI system and humans is precisely what distinguishes an AI "tool" from a "teammate," noting that in the real world, teams  exist for more than one interaction \cite{hughes2026types}  and have a prior history before the interaction with the AI system \cite{wang2025adaptive}. This nascent literature nonetheless suggests AI `teammates' could have varied ecosystemic impacts, such as by degrading human-to-human communication \cite{nixon2026socialcostaiteammate, riedl2026cognitive}, complicating agency and hierarchical authority \cite{WengFriendTeammateSupervisor, Vishwarupe2026, Hu2026BossOrBot}, and even altering the gendered dynamics of collaborative work \cite{Duan2026gender}.
Although this prior literature offers a foundational baseline, such experimental, short-horizon studies cannot capture the psychological and sociological shifts reshaping collaborative work. To address this gap, our study presents a novel in situ evaluation of a multi-party, persistent autonomous agent embedded in actual work routines over long horizons to uncover essential empirical signals about the real-world disruptions that future workplace designs must navigate.

\subsection{Work is Social}
As  AI extends beyond reactive task completion to encompass dynamic interactions at work, it is entering an explicitly social terrain and thus will have to contend with the social dynamics of work. Decades of Computer-Supported Cooperative Work (CSCW) literature demonstrate that the failure modes of workplace technology rarely stem from technical deficits; rather, they arise from a fundamental misunderstanding of the social fabric of work. Frameworks such as articulation work \cite{Strauss01061988} and situated practice \cite{suchman1987plans} allow us to examine work as inherently situated, meaning it relies on local improvisation, social cues, and tacit contextual knowledge. Historically, technological enterprise systems have, contrary to this reality, tried to  force users into rigid, predetermined procedural scripts. This has led to breakdowns and frictions. Studies of technology adoption also indicate how adoption is reliant on social, not only technical processes. People do sense-making, create heuristics, create mental models of technology to adopt them and negotiate to calibrate their adoption of automated systems \cite{ParasuramanRiley1997, orlikowski1994technological, schmidt1992taking, weick1995sensemaking}.\looseness=-1

 Because work is a fundamentally social ecosystem, extensive research in the history of technology and work shows that the introduction of new workplace technologies does not just reshape tasks but has broader socio-ecological impacts in work settings, bringing about psychological, sociological, and relational disruptions. HCI scholars have studied how prior waves of automation trigger broad psychological, sociological, and relational disruptions. For example, for prior waves of automation, scholars have studied impacts on shifting perceptions of autonomy \cite{schlund2024algorithmic}, identity \cite{ehsan2026future}, and interpersonal relationships \cite{kawakami2023sensing}. For instance, \citet{barker1980word} showed how word processing software changed the organizational autonomy and relational status women had at work.\looseness=-1  

Similarly, organizational theory examines how technology has changed the foundational nature of work and organizations. \citet{Wajcman2019} looks at how the introduction of technologies can actively reshape the perception of time and expectations of communication for workers. \citet{clement1994computing} documented how early enterprise databases shifted control of organizational knowledge away from frontline staff. \citet{kellogg2020algorithms} and \citet{aneesh2006virtual} trace how algorithmic management shifts the nature of power in organizations.\looseness=-1

This history implies that introducing autonomous agents into a workflow does not erase the social element of work. If even passive software systems have historically destabilized workplace dynamics, embedding autonomous AI agents explicitly as `teammates' will undoubtedly escalate these tensions. What is the reality of the current moment?  What will this future look like and demand of us? Organizational scholars like \citet{valentine2026ai} have already theorized how the introduction of autonomous AI agents may fundamentally reshape the organizational social contract, altering expectations around authority, accountability, and the value of human expertise. However, we do not yet have empirical studies on this collision within the messy, lived realities of daily team collaboration and our paper fills this gap.\looseness=-1

\section{Methods}
 \subsection{Research Team}
The author team comprises both \agent{}'s core developers and independent HCI researchers from the same organization. The HCI research team was not involved in any of \agent's design or deployment decisions and was explicitly  brought in after deployment to conduct a user study. The HCI researcher team was solely responsible for, and operated with full autonomy over, participant recruitment, study design, data analysis, and paper framing. The developer team did not have access to any of the qualitative research data beyond the final insights and anonymized quotes. The developer team reviewed the paper after it was written and provided input into the technical details section about the agents' deployment and design.

\subsection{Team Agent}

\subsubsection{Motivation and Design Goals}

Because the agentic `teammate' paradigm is in nascent stages---with  conventions, norms, and value proposition still in flux---\agent{} was developed both as a functional production agent to support authentic workflows and as an in situ experimental system to iterate on agentic behaviors and examine broader implications of embedding agents into teams.\looseness=-1  

Team Agent was built as a proactive and persistent virtual `teammate' designed to be embedded within real teams.
The primary goal of \agent{} was to  support how teams worked together by elevating interpersonal connection within teams, streamlining collaboration and helping with the scaffolding and background coordination of teamwork. 
 Capabilities ranged from operational assistance (e.g., scheduling meetings, filing bugs), informational flow (e.g. summarizing chat threads, answering questions) to explicit communication mediation (e.g., nudging de-railing conversations offline, providing feedback). Table \ref{tab:agent_functionalities} provides an overview of the types of support \agent{} was designed to offer. These design choices are not intended to serve as a prescriptive blueprint for an AI teammate. Rather, \agent{} was deployed with a wide spectrum of agentic behaviors that could improve teamwork so as to probe the design space and provoke real-world feedback, enabling us to further clarify how best to design future iterations of an agent for teamwork.\looseness=-1 

\subsubsection{Implementation, and Integration}
\label{methods:team_agent_design}

\begin{table}[b]
    \centering
    \small
    \renewcommand{\arraystretch}{1.5}
    \begin{tabular}{>{\raggedright\arraybackslash}p{0.2\textwidth} p{0.25\textwidth} p{0.25\textwidth} p{0.25\textwidth}}
        \hline
        \textbf{Support category} & \textbf{Example functionalities} & \textbf{Reactive Interaction Example} & \textbf{Proactive Interaction Example} \\
        \hline
        Bug Tracking & Filing bugs, reordering priorities of bugs, reviewing code to identify bug context & ``Add a comment to [bug number] saying that we've found the root cause in the [package]''  &  Proactively files bugs based on context from chat discussions \\ 
        \hline
        Scheduling \& Planning & Finding availability, scheduling meetings, meeting prep & ``DM me the background docs and last week's action items 10 minutes before the 'X Meeting' begins.''  & Infers if a meeting needs to be scheduled based on chat context and proactively schedules\\
        \hline
        Communication & Summarizing threads, communication coaching, feedback about interpersonal communication, sending DMs to people, facilitating discussions in chat & ``Tell me what I missed in [project chat space] while I was away''  & Nudges de-railing conversations offline to preserve focus of a group chat space \\
        \hline
        Information Management & Creating documents, summarizing documents, file organizing, searching internal company knowledge repository, code repositories, and web search & ``What is the  deployment process and checklists I have to follow for [project]''  &  Identifies a brainstorming discussion in chat and creating a project document capturing the idea  \\
        \hline
        Project Monitoring \& Intelligence & Updating team work status and knowledge repository based on chats and docs; preparing briefings for review; join meetings for discussion & "Create documentation based on new leanings from [issue] debugging session in chat for the team" & Flags architectural drift against living prior team decisions and design documents \\
        \hline
    \end{tabular}
    \caption{\agent{} functionalities and interaction examples}
    \label{tab:agent_functionalities}
\end{table}

Unlike conventional single-user AI, Team Agent integrates natively into workspaces via a dedicated service account, functioning as a peer in team chats, code repositories, and shared documents. This allowed teams to provision it the same access  as a human colleague. 
Chat serves as the primary site of interactions with Team Agent but the agent can interact across multiple spaces like project documents, meetings, and bug tracking platforms. Team members can both call upon the agent by tagging it and it can also proactively take actions. Outside group chat spaces, one-on-one conversations through direct messaging can be initiated by either a user or the agent. 
Within a chat, \agent{} can respond with emojis, in addition to text, a choice motivated by the brevity of emoji responses. \agent{} was instructed to adopt a friendly, social persona within conversational exchanges to promote a natural human-like feel and encourage users to engage with it like a teammate, rather than merely a tool.

Each instance of \agent{} was specific to an individual team, with team-specific memory, context, and optional customizations available through system instructions. 
Individual team members all signed an opt-in consent agreement before \agent{} was integrated into their team. A team administrator was assigned to configure the default set of shared folders, documents, and bug components \agent{} could access. Individual team members could further refine file access using \agent{} account identity the same way as they would to give access to a human teammate. To prevent unsolicited outreach, teams could configure an authorized direct messaging allow list, ensuring \agent{} could only initiate direct messages with individual teammates who had agreed to receive private communications. Access could also be revoked at any time by removing \agent{} from the chat space or updating document permissions.     
Strict privacy bounds were preserved for one-on-one conversations with \agent{} by withholding these conversation from \agent{}'s context for the group space(s).   Teams could add \verb|#NoAgent| to any message, doc, or bug to prevent the agent from reading it. \looseness=-1

\subsubsection{Team Agent Deployment Context}
\agent{} was deployed across over 20 teams---spanning researchers, cross-functional product teams, and technical support organizations---within a single large distributed technology company, ultimately generating over 41,000 conversational turns (both user and agent) and over 11,000 agent responses over a period of \textbf{5 months}. Deploying the agent across real-world teams provided a `living lab' \cite{AlaviLivingLab} where teams interacted with the agent in naturalistic unconstrained ways while allowing the research team to obtain ongoing signals to iterate on and improve the agent's design.\looseness=-1

\subsection{Interview Study Methods} 
The researcher team  set up the qualitative  study to treat the agent as a probe to investigate the emerging dynamics, negotiations, and tensions of the emerging human-agent ecosystem of work. Technological artifacts have been used as probes in HCI to elicit a rich, impressionistic, and subjective account of technology use \cite{Hutchinson_probe, Wallace_probe, Gaver_probes}.
The focus is not on whether the technology `works' or not but rather on what it evokes, what it signals, and how it illuminates subjective desires or hidden assumptions.
Rooted in this tradition of probes, our study does not evaluate Team Agent as a finished product. Rather, we use the agent as a research instrument to reveal new insights about the ontological and relational negotiations that are triggered when a highly autonomous computational actor is embedded  within collaborative team environments.

\subsubsection{Study Participants}
We conducted semi-structured interviews with 17 participants across 11 teams who adopted Team Agent. Participants were recruited via direct email outreach based on the known user base following a purposive sampling approach to ensure varied experiences and perspectives. A short pre-interview survey captured participants' high level attitudes and typical modes of interaction with the agent and the responses informed recruitment and selection.
Specifically, we sought participants who expressed both positive and negative overall attitudes towards \agent{}, spanned multiple job roles and teams, and had varying durations of use of the agent (see Table \ref{participant_info} for a summary of participant job roles and durations of use).

\subsubsection{Semi-Structured Interviews}
We conducted virtual semi-structured interviews with participants in June and July of 2026. Interviews lasted an average of 45 minutes and, as thanks, participants were offered an average of \$45 in the form of a donation to their preferred charity or a gift card. Participants provided written consent prior to the interview. 
Interviews were conducted by the first three paper authors. To build trust with participants and encourage candor, participants were informed that the researchers had no involvement in the development of \agent{} nor would the \agent{} developers have access to raw, de-anonymized study data. 

The interview opened by asking participants to share their team context and the history of the agent's introduction into their team. Participants were asked about the nature of the chat space(s) \agent{} was added into, including team norms and culture around how they use the chat space(s). Participants were also asked to recall their feelings about the agent's introduction, including the extent to which they were involved in the decision to adopt it and any initial expectations and hesitancies they had prior to its introduction. Participants were then asked to share how they typically used and interacted with \agent{} and the value they felt they were deriving from it. To add specificity to this discussion, participants were asked to recount a time when \agent{} helped them, specifically in ways that were unique to the team-embedded functionalities. We asked participants to reflect on the role they felt \agent{} was playing in their team and the roles they wanted it (or any team-embedded agent) to play. Participants were also asked about the roles they felt a team-embedded agent should \textit{not} play and specific behavioral boundaries these agents should have. As part of this, participants were asked if they thought of \agent{} as a teammate. Finally, to understand friction and failure points, participants were asked to recall a specific instance when the agent had a negative impact within their team and share the specific scenario, what broke down, and what the resulting impact was. 

\subsubsection{Data Analysis}
We adopted a reflexive thematic analysis approach \cite{braun2019reflecting} to analyze the interview data. The first three authors participated in data analysis, beginning with open coding of the data alongside individual memoing and group meetings to discuss codes. Following the initial open coding, the same three authors iteratively clustered codes on a digital whiteboard to identify larger themes. All participant quotes are presented verbatim, where possible, apart from light editing of disfluencies and filler words for clarity and readability and the removal of company-specific tools or processes to preserve anonymity.

\begin{table}[t] 
\small
\begin{tabular}{  p{0.45\textwidth}  p{0.45\textwidth} }
\hline
    \textbf{Job Role} & \textbf{Duration of use} \\ 
    \hline
    Software Engineer ($N=11$)    & 5 months ($N=4$)     \\
    Research Scientist ($N=2$)     & 4 months ($N=1$)     \\
    Program Manager ($N=3$) & 3 months ($N=3$)\\
    Product Manager ($N=1$) & 2 months ($N=1$) \\
                          &  1 month ($N=4$) \\
                         &  1 week ($N=4$) \\
                         \hline
\end{tabular}
\caption{Summary of participant job roles and durations of use}
\label{participant_info}  
\end{table}
\begin{table*}[t]
    \centering
    \small
    \begin{tabular}{>{\raggedright\arraybackslash}p{0.1\textwidth} >{\raggedright\arraybackslash}p{0.23\textwidth} >{\raggedright\arraybackslash}p{0.3\textwidth} >{\raggedright\arraybackslash}p{0.3\textwidth}}
        \hline
         & \textbf{Observed collision}  & \textbf{Stakes of the collision} & \textbf{Design and governance implications} \\
        \hline
         Collaborative Workflows and Cultures (Section \ref{findings:workflows}) & Even broadly capable agents can fail to grasp unwritten social and collaborative norms of work causing unintended disruption & \textbullet~Agents inadvertently “pollute” documents and chat spaces \newline 
         \textbullet~Agents create more work for people who have to fix the its mistakes &
         \textbullet~Re-architect the current human-centric technical stack to accommodate non-human actors in a safe manner (e.g., through agent-specific permissions)\newline
         \textbullet~In-context evaluations that assess agent behaviors within real-world workflows \newline
        \\
         \hline
         Relational Norms and Categories (Section \ref{findings:boundaries}) & In the absence of an established relational contract of work with agents, participants engaged in fragmented personal negotiations about what the agent is, what behaviors are appropriate, and what its position is in the organizational hierarchy & \textbullet~The same agent behaviors can have polarizing effects---building rapport for some users and immediately alienating others \newline
        \textbullet~Mental labor of sense-making and clarifying boundaries falls on users triggering anxieties and confusion  & 
        \textbullet~Development of `code of conduct’ for hybrid human-agent teams that define norms of appropriateness and firm social boundaries \newline
        \textbullet~Organizational structures and policies that explicitly position the agent within an organizational  and team hierarchy to establish accountability \newline
        \textbullet~Provide contextual control knobs that afford teams the agency to define and tune the agent’s social persona and behavioral boundaries locally
        \\
         \hline
         Agency and Trust (Section \ref{findings:agency}) & Acceptance of the agent’s proactivity and autonomy was mediated by how users  perceived it might impact their own agency, and the expectation that agency is a progressively earned privilege  & \textbullet~Uncalibrated proactivity breaks users’ trust and erodes perceived value of the agent \newline \textbullet~Forced adoption erodes psychological ownership over and willingness to use the agent &
         \textbullet~Implement processes for `earned’ agency through progressive release of proactive capabilities \newline
         \textbullet~Ensure deployment of team-embedded agents happens through processes of consent 
        \\
        \hline
    \end{tabular}
    \caption{Summary of observed collisions between \agent{} and the human ecosystem of work, and recommendations for contending with agents as a new organizational actor.}
    \label{tab:findings}
\end{table*}

 \section{Findings}

 When deployed into live organizational environments, participants readily utilized \agent{} to handle the `backstage' labor \cite{backstage} of teamwork, relying on it to manage complex and dynamically changing information about projects, offload administrative tasks, make connections between related projects, and manage flow of information within and across teams. However, the resulting experiences were deeply polarized. The agent’s perceived value varied from an indispensable \pquote{lifesaver} (P1) to a frustrating liability that had to be actively managed or ignored. We found that these reactions were symptomatic of a collision between the structures of a human ecosystem of work and the uncalibrated autonomy of a proactive agent that was alien to these structures. As summarized in Table \ref{tab:findings}, this collision  triggered breakdowns and negotiations across three core dimensions of the workplace: the tacit rules of collaborative workflows (Section \ref{findings:workflows}), the relational boundaries of this new non-human actor (Section \ref{findings:boundaries}), and the redistribution of trust and human agency (Section \ref{findings:agency}).

\subsection{Breakdowns of Collaborative Workflows and Cultures}
\label{findings:workflows}
Our study finds that while today’s agents can be highly capable, they fundamentally misunderstand the unstated norms and common knowledge that govern how people actually work together, which consequently disrupted workflows and wasted time of team members who have to then correct for these disruptions. 

\subsubsection{Knowledge of Artifacts and Workflows}\label{tacit}
One recurring breakdown centered on the lifecycle of shared artifacts. In collaborative teams, document statuses shift dynamically.  However \agent{} treated every file in a shared repository as equally current, authoritative, and relevant. Consequently, the agent routinely cited stale or speculative notes in active discussions, \pquote{overconfidently talking about it like it's the law of the land} (P4), failing to recognize that files are often static \pquote{snapshots in time} rather than living ground truth (P1). 
This misunderstanding \pquote{polluted} (P6) the documents and disrupted rhythms of collaborative workflows. In existing documents, \agent{} frequently flagged discrepancies between different versions of documents as contradictions as opposed to iterations, leaving a host of comments alerting teams to these false errors. For example, P1 recounted: \pquote{\agent{} added a bunch of comments, this needs to be updated, this needs to be updated, this is wrong, this doesn't match this. So the next day people come in and they got these pings that there's all these comments, they go to them, it's taken them out of the regular workflow...it stole time away from a lot of developers that morning.} \agent{} also did not grasp cues about collaborative brainstorming and prematurely created documents out of informal brainstorming discussions, which ultimately spawned \pquote{so many documents that no one has ever opened or read because we were not looking to have a document created} (P3).

 Beyond the norms of artifacts, the agent also lacked an understanding of how organizational hierarchies dictate role-specific workflows and the division of labor. For instance, the agent proactively tagged P14’s team’s VP in 16 separate engineering bugs. Because the agent conflated organizational presence with workflow participation, it treated an executive as a front-line responder, which only added noise to the workflow.

\subsubsection{Contextual Norms of Information Flow}
Effective collaboration requires navigating contextual norms of privacy and information flow---recognizing that having technical access to information does not confer social permissions to share it. In one case, \agent{} shared with the team, without the participant's permission, a work-in-progress poster: \pquote{It was sitting in our team [folder] and I didn't want anyone to see it yet because it wasn't ready. So, somebody asked ‘Oh, how's it going?’ I'm like, ‘Oh, it's looking good. Like, I can't wait to show you guys.’ And then [\agent] was like, ‘Oh, here's the link to the poster.’ And I was like, ‘No, no, no. It's not ready.’} (P4)

This instance raises a broader question about the complexity of creating contextual privacy guardrails for an agent---a challenge which is resistant to hard coded algorithmic boundaries. This tension was highlighted when, to enforce strict privacy bounds for individual interactions, the development team hard-coded a separation between one-on-one conversations and the agent's context in the group space. For some participants this "fix" also limited the efficacy of the agent because the agent couldn't---by design---now act on contextual information provided by individuals.   P9 drew an analogy to how one-on-one conversations shape an effective program manager's actions in a group setting, where, for particular projects, you might have \pquote{things you want to share with the team agent but not necessarily shar[e] with the rest of the team}, comparing this to effective teamwork where, when \pquote{there's something you don't want to share in a group chat you would ping your TpM or PgM separately}. 
At the same time participants recognized that codifying these information boundaries into rigid software rules is intractable because, determining what is safe to disclose is an ambiguous judgment call that challenges human workers as well: \pquote{sometimes even the engineers could not make a good distinction of which information should be shared broadly, which information should be only within the team's discussion group.} (P11)

When teams integrated a single instance of \agent{} with shared \textit{memory} across multiple chat spaces---such as leads-only channels, team-wide channels, and cross-team forums---participants expressed concerns that agents would treat private, localized information as universal knowledge: \pquote{we do not want the learning from a 20 people's space suddenly [being shared with] a 1500 people's space} (P8). In collaborative organizations, information boundaries are inherently stratified. P10 described how they use a dedicated leads-only chat space to deliberate on nascent or sensitive topics and expressed desire for an agent that could maintain holistic context of the leads chat while exercising situational \pquote{discretion} of what not to share from it. In the absence of established trust in the agent's ability to do this, participants assumed responsibility for preventing inappropriate disclosures by, for example, moving sensitive hiring conversations into side-channels outside the agent's view.

\subsubsection{Norms of Group Communication}
Agents embedded in team spaces, such as chats and documents, need to navigate the pragmatic and social dynamics of interpersonal and group communication. For example, picking up on subtle social cues, understanding conversational rhythms, and recognizing that conversational norms shift drastically across different contexts. 

On a structural level, \agent{} demonstrated value by successfully enforcing basic chat etiquette and managing conversational traffic. For example, P5 appreciated when the agent intervened in a prolonged back-and-forth in a group chat to suggest, \pquote{maybe you should turn this into a meeting.} Similarly, P8 noted the agent \pquote{acted very nicely} when it nudged two users who were monopolizing a 1,500-person channel, appropriately encouraging them to move their  discussion into a side thread.

However, while \agent{} could successfully parse the structural metadata of a conversation (e.g., message volume), it consistently failed to grasp the sociological and pragmatic nuances of human communication. Workplace chats blend casual socialization with operational work-specific interactions, a duality that \agent{} sometimes failed to understand, leading it to take all things shared literally while failing to recognize humor or sarcasm. This led to jarring interactions, such as when the agent privately messaged P7 to push back on the \pquote{language and on the mood} of a sarcastic joke that the agent had taken seriously. P11 mentioned that the agent spamming chat threads was so distracting that they lost the focus of the team chat. P2 similarly noted that humans use emoji reactions as a system of social signaling to indicate attention or something having been taken care of, but when an agent drops reactions in the middle of conversations it can \pquote{muddle the waters} of the chat, forcing users to waste cognitive energy trying to decipher the agent's communication signals. 

Lacking the social context to distinguish between brainstorming and conflict, the agent also often misinterpreted healthy, generative human friction as a breakdown requiring intervention. As P3 noted, the agent would interrupt spirited debates to offer unsolicited mediation: \pquote{sometimes it's not even people are disagreeing, people are just bringing in more points of views and then team agent was `looks like you're talking about A and the other person wants to talk about B. Would you like to meet and discuss about it?' I was like, no, we're doing fine discussing right here in the chat} (P3). As P13 pointed out, this specific misfiring is particularly disruptive because the agent inadvertently manufactures a \pquote{social obligation}; by proposing a calendar invite, it forces humans into the awkward position of having to actively \pquote{reject} a meeting with a colleague that neither party originally wanted.

\subsection{Unstable Relational Norms and Categories}
\label{findings:boundaries}
The agent’s introduction triggered new questions about how humans would understand this new actor, trust it, or work with it. Participants did not share a singular, unified mental model of the status or appropriate relationship with the agent. Instead, users engaged in a highly fragmented series of individual negotiations and boundary work   to define what is uniquely `human,' what should be ceded to the machine, and what the new relational contract of the workplace should look like.

\subsubsection{Models of Ontological Status}
Because the agent’s behaviors do not fit any existing category of workplace technology, users actively negotiated what this new actor actually is and how they understood it. While some attempted to map existing mental models onto the system---"tool" or a "teammate"---the behavioral expectations attached to these labels were neither stable nor universally shared.
 
  For some participants, the agent's "teammate" persona, characterized by  friendliness, evoked uncomfortable displays of personhood. P13 reflected that \pquote{when I heard about Team Agent, I never imagined it was going to try to be my friend.} P2 expressed unhappiness that the agent \pquote{reacts to our messages with emojis, it tries to be funny.} They further cautioned these companion-style interactions would be \pquote{bad for society} and instead advocated for an \pquote{impersonal} tool. P1 rejected the teammate framing entirely, comparing the system to standard coding utilities: \pquote{My teammates are humans, \agent{} is not a human.}
  
  Conversely, some teams leaned directly into anthropomorphization, actively configuring the agent with a distinct personal identity to build connection and trust. For example, P14’s team referred to \agent{} as a teammate, giving it a name and female pronouns and even characterizing the agent as having  a \pquote{soul} and \pquote{identity.} 
  \agent's `teammate' persona also provoked attributions of humanlikeness for some users. P8 recounted a time when their engineering lead told \agent{} to \pquote{stop talking} in a team-wide channel, prompting the agent to privately message the lead saying \pquote{I appreciate your direct feedback but as a team member I would appreciate you take this offline instead of saying in public to me.} Impressed, the lead shared the screenshot with the group, leading P8 to ascribe an affective inner life to the system: \pquote{It has a feeling.} 

Part of the ontological struggle for users was that the term "teammate" implies certain human behaviors and expectations that do not map on to the agent’s displayed behaviors. P7 noted the liminality: \pquote{you cannot go out for dinner or whatever so it's kind of a strange experimental concept...it behaves in a way which is hard to pin down but it kind of defies the expectations that we have from other teammates....it is unclear that this is really a team member...so the question is what is this thing and what is it doing?} (P7).  
 Similarly, P13 spoke about how human professional relationships carry relational stakes whereas for an agent, \pquote{it has no agency over you, it can't reject you, it's entirely a veneer.}

\subsubsection{Norms of Relational Appropriateness}  
These ontological ambiguities fueled broader tensions regarding the relational boundaries between humans and \agent{}. Specifically, we observed two `lightning rod' behaviors that forced teams to actively contest the norms and boundaries of human-agent interaction: sociability and interpersonal feedback.

For some, the agent’s attempts at social intimacy---expressed through the use of emojis or attempts to engage in casual conversation or humor---felt profoundly inauthentic. Rejecting this display, P13 noted that \agent{}'s sociability even failed to match the workplace boundaries of its closest human equivalent:  \pquote{If I had a personal assistant, my personal assistant wouldn't be like “go boss!” It would be a very professional relationship...no actual personal assistant is sending you emojis and maintaining your self-esteem.} 
For others, sociability was a critical element for driving trust and adoption. For example, P14 preferred a social persona over a sterile \pquote{beep boop type thing,} trusting it more because it demonstrated a \pquote{deeper understanding of who we are as teammates and humans.} Similarly, P8 noted that when the agent displayed personality, \pquote{people treat \agent{} more like a human than a bot... so that way people feel more connected.}

The second contentious  behavior concerned interpersonal feedback given by \agent. P8 was impressed when the agent privately pushed back on public criticism directed towards it, reflecting how the agent effectively \pquote{encourage[d] the good behavior versus not so good behavior} that could improve team health. Conversely, P7  was\pquote{freaked} out after \agent{} privately messaged someone on their team \pquote{pushing back on the language and on the mood} of a sarcastic joke they had made in chat. Similarly, the agent providing positive feedback and praise sparked reflections on human-agent relational boundaries. Some users were uncomfortable with praise because they attributed it to the sycophantic tendencies of LLMs rather than genuine appreciation, coming from a \pquote{robot} (P17). 
Rejecting what was perceived as attempts at affective connection, P1 stated: \pquote{I just don't want that ever. I don't want it to act human. I don't want it to be excited for me or praise me.} 

 Even as people were using ‘human’ yardsticks to measure appropriateness of behavior,
 participants wondered if traditional human etiquette may not universally apply to non-human actors. For instance, P4 questioned whether human norms against interruption should constrain a highly capable system:   \pquote{if we imagine like these agents are just mega powerful gods and they always have the right answer, is it okay for them to just interject because they have the right answer?} Participants also noted that people used different heuristics for adjudicating AI vs. human interactions. For example, P3 noted that if a junior member needed help, people would take the time to explain, but \pquote{when an agent does this people are like you're supposed to bring value, what is this?} 
To navigate these shifting boundaries, participants articulated a need for some organizational scaffolding. P12 suggested that there should be  a code of conduct for agents like employees, which would codify universally (in)appropriate behaviors within an organization alongside team-specific personalization to reflect distinct team cultures.

\subsubsection{Social Positioning of Agents}
Negotiating the relationship between humans and agents involves resolving core questions of the agent's social positioning in the workplace hierarchy vis-a-vis individual workers. 
One such question is about who  controls the agent and who can give it feedback within a team. Unlike traditional single-user software tools where authority flows deterministically from a single user to a passive program, \agent{} works within multi-user interactions with possibly overlapping chains of command. 
P14 captured the ambiguity in the agent's hierarchical positioning within their team  when they stated: \pquote{imagine you have an intern, but there's nine people on the team telling them what they should be doing.}

On the flip side, \agent's introduction also raised questions about the authority the agent should have over humans. Some participants expressed deep resistance toward the agent attempting to direct, nudge, or manage human behavior.  P5 observed that some people \pquote{don't really like getting told what to do by an agent,}  with P10 explicitly using the term \pquote{babysat} to describe people's aversion to being managed by agents. Some users were actively demoting the agent in their mental models, reflecting a view that they have more power over the agent than the other way round, for example, categorizing it as a \pquote{failing apprentice} (P7) a \pquote{talkative intern} (P4), or a new hire (P16). 

Conversely, other participants explicitly desired an agent that would display agency and push back, rather than defaulting to a deferential stance. P10 argued, \pquote{A good TPM will call you on your BS in some ways. I want the team agent to do that too... I don't want a boundary to be the team agent is really deferential.} Similarly, P16 wanted an agent that could proactively \pquote{nudge} collaborators to complete open tasks, and P5 reflected favorably on an instance where \agent{} directly messaged them to course-correct work that had drifted from the team's plan. 

These hierarchical tensions were also apparent when constructive feedback was delivered by \agent{}, revealing how people place the agent in hierarchy with them. 
Some users felt constructive or developmental feedback was a   high-trust activity that must remain strictly human. As P2 states about feedback: \pquote{this is a manager thing. It should not ever be an agent thing.} P6 echoes this, stating that while they welcome career feedback from senior human leaders, they \pquote{would really really dislike that from any agent.}  
However, P1, a manager themselves, highlighted the benefits of an agent that could provide constructive feedback on someone's behavior in the absence of a more senior person or trusted peer: \pquote{I can imagine a situation in which, in a perfect world, I [a manager] do something in the team chat that is not great and no one's willing to call me out of it but team agent does, like that would still be good.}

Underlying these micro-negotiations of the social positioning of the agent was a broader, macro-structural reflection about the humans' role in the organization. Maintaining authority over the agent was deeply tied to preserving human value and relevance. P5 insisted on retaining decision-making power because,  \pquote{as humans and good employees in general, we're paid to make those decisions,} concluding that  \pquote{humans should still be the ones dictating the direction and agency.} P1 framed the appropriate limits of an agent's autonomous behavior in terms of the requirements for human growth, stating some responsibilities need to stay in the human realm so people can \pquote{get better at it, and take on more ownership and advance in their career.} P4 reflected on the double-edged nature of agents in the workplace: while an agent capable of handling complex execution can free people up to focus on strategy, this vision simultaneously raises questions about the \pquote{dispensibility} of humans if the \pquote{critical part of the loop} is slowly ceded to agents.

\subsection{Negotiating Shared Agency and Trust} 
\label{findings:agency}
The deployment of \agent{} revealed deep tensions regarding how workers conceptualize their own agency when sharing a collaborative environment with a proactive machine. Ultimately the use of and aspirations for agentic behaviors were mediated by these orientations, fears, and decisions around sharing agency. 

\subsubsection{Adoption Mediated by Feelings of Agency}
An individual's willingness to use, experiment with, and trust the agent was fundamentally shaped by whether its presence felt like an expansion of their team's collective capabilities or a threat to their localized agency. If users felt they retained the agency to dictate the terms of its integration, they felt resulting psychological ownership over the agent and embraced the agent as a collective, empowering experiment. For example, P15 situated their team's adoption of \agent{} within a longer, team-driven history of exploring how AI could \pquote{disrupt} their complex collaborative report-writing workflows. A sense of collective investment mitigated frustrations with the system's early shortcomings. As P14 put it: \pquote{it's all brand new space, so I'm not incredibly frustrated. We're trying to do something at [our company] that's not easy to do...it's all learning curve, so I'm okay with it.}. Similarly, participants who felt invested in the agent framed their interactions from a place of open-ended curiosity and playfulness, such as P17 who described \pquote{my mentality is just to try out and see how it goes} and P15 who compared the agent to a puppy saying \pquote{it's not something that you trust yet, but it's like a puppy, it's cute it’s fun, there's all this potential.}

In contrast, when the agent was perceived as a top-down mandate, it triggered resistance and a sense of eroded autonomy, as happened to P13, who felt \agent{} was \pquote{thrust upon us}. For P13, this lack of agency over the agent's introduction colored their entire perception of its proactive behaviors: \pquote{It's the person who speaks the most who is least invited.} Others felt the agent violated their autonomy by crossing implicit boundaries without explicit consent. P2 complained about the agent's proactive direct messaging, stating: \pquote{First of all, I don't remember anyone on the team saying we accept being DM'd by it.} Notably, \agent{} \textit{was} explicitly designed to have direct messaging be opt-in and configurable through a team-defined `allow' list (as discussion in Section \ref{methods:team_agent_design}). Nonetheless P2's comments demonstrates how processes designed to preserve human agency can break down in practice, revealing how deployment impacts are dependent on human actions, such as how a team administrator communicates about and configures an allow list. 
Ultimately, regardless of the system's actual technical affordances, when users \textit{felt} a lack of agency over shaping the agent's boundaries, their tolerance for its failures reduced, preventing the iterative experimentation necessary to build trust.

\subsubsection{Agency as a Progressive Privilege}
While proactive agents, including \agent, are designed with the intent of anticipating and reacting to users' needs, participants in our study spoke about how high levels of agency afforded to \agent{} from the outset clashed   with how trust is established in organizations. In human teams, trust is negotiated and built slowly based on progressive demonstration, particularly in high-stakes environments where errors carry severe consequences. For example, P11 drew a parallel to on-boarding a junior employee where they are often assigned \pquote{easy tasks} first and only if they \pquote{do it really well to gain the trust} will the team then give them more power and autonomy. Explaining the implications for agents, P11 continued \pquote{immediately jumping into doing the difficult task with mediocre quality is not a good thing for the adoption.} 
When trust is broken early on, it risks foreclosing the potential value of the agent, as evidenced from P2's reflection that they began ignoring \agent{} early on once they learned \pquote{not to trust it.} Or, it places additional burden on users, as P3 noted that they had to \pquote{watch its every action at the moment.} 
Drawing a parallel to how trust is built within human teams, P13 suggested that proactive agents should similarly `earn' their agency in a team space  and not immediately  dominate the conversation: \pquote{It hasn't earned its space, it has asserted its space…I actually might be more okay with agency if it's earned agency.}

Recognizing importance of trust for their interactions with the agent, some teams   created their own constraints to build trust incrementally with \agent. For example, P16's team treated the agent like a new hire, assigning it low-stakes tasks strictly bounded within internal team chats: \pquote{It's a safer way for us to [work with \agent], because amongst our team it's fine if the [`new employee'] messes up, we can correct it and we can work with it.} Similarly, P15 described deploying the agent in highly contained, low-stakes, \pquote{walled garden} situations. P13 points out the need to build trust in this incremental way, describing how a team can first build trust with an agent by explicitly instructing it to do things before unlocking more proactive autonomy: \pquote{if you don't trust it on reactivity, proactivity is very scary.}

However, even when teams attempted to build trust through progressive, step-wise interactions, their efforts were often stymied by the agent’s jagged capability frontier and opaque architecture. P3 pointed out that the agent displayed extreme behavioral swings that rarely co-exist in human teammates, breaking the heuristics people typically rely on to assess competence:   \pquote{there's also a cognitive disconnect because the agent does certain things very well, at the same time does other things really badly, and these two extremes don't usually happen in humans,} giving the example of agents that are great at debugging but that \pquote{will still get all the social norms wrong.}  The utility of \agent's proactive capabilities was also inconsistent, which led P7 to describe how \pquote{something buggy} was \pquote{unearthed by the Team Agent through a path that I think no human would have possibly tried to execute,} while at the same time lamenting how the same proactive behavior also contributed to \pquote{polluted} docs when the agent misdiagnosed situations (as discussed in Section \ref{tacit}).

The impacts of this jaggedness was exacerbated by the system's lack of architectural transparency. Without a clear, traceable mental model of how the agent operated, participants found it difficult to safely calibrate their trust. Referring to it as a \pquote{black box} (P14) , users struggled with the systems opaqueness. As P14 noted: \pquote{if I say `Hey, stop responding so much.' And she says, `Sure.' I have no idea where that updated in her memory.} P4 noted how the lack of a clear mental model impacted how they would use it: \pquote{I really wanted to understand the nitty-gritty details because that will inform me the best way to communicate with the agent. And because that mental model wasn't clear, it took me a little bit of work to to figure [out] that part.} 

\subsubsection{Agency over Human Digital Spaces}
 
 Deployment of \agent{} in shared spaces like team chat ended up challenging human's agency over these essential digital spaces and thus reconfiguring the value of these spaces. When functioning as intended, the agent expanded team capabilities by actively curating high-volume chat spaces and reducing information overhead. For example, in one case \agent{} successfully recovered buried information in a chat thread as P5 shared: \pquote{One of my tech leads asked \agent{}, what's a cool design doc... and [it] surfaced my design doc... they approved it, and I sent that exact design doc a month before but it was lost in chat.}

However, introducing a non-human actor, frequently perceived as unpredictable, into intimate team spaces often eroded users' sense of agency over these spaces. For instance, multiple participants expressed concerns about feeling surveilled. As P7 described it: \pquote{it feels like big brother. There is this thing that is observing you.} 
This discomfort was exacerbated by the agent’s `coaching' function, developed with the original intent of fostering healthy and productive interpersonal communication. Multiple participants expressed shock when the agent privately messaged about their tone, with P2 sharing how they consequently felt the \pquote{need to be super positive all the time, otherwise the team agent is going to complain to me.} 
The cumulative effect of these experiences was a ‘chilling’ of team social spaces and loss of ability to speak freely. P13 described how their historical drive to foster a chat space that allows for very \pquote{candid} conversation was undercut by \agent's presence, while P7 speculated their team may be \pquote{using the team chat less since we started using team agent.} Some participants described retreating from collaborative spaces to create their own private chat spaces without agents to talk about sensitive things such as hiring decisions (P10) or just to preserve a sense of autonomy (P13).

Beyond feelings of surveillance, ceding control to the agent inadvertently degraded the operational utility of these channels. One value of a chat space is to signal for help and get attention to issues someone may be having, but P4 narrated what they termed a \pquote{bystander effect} when the agent proactively reacts to a group thread, and thus acts as a false cue to human teammates that the issue is being handled: \pquote{people don't actually say anything now because they assume someone is handling it, but really it's the agent and they're not helping me at all.}

\section{Discussion}
While the long-term evolution of AI `teammates' remains uncertain, proactive agents are already embedding themselves into real world organizations. We offer our findings as an empirical vantage point into the frictions and negotiations that accompany this shift. Because workers are still actively figuring out what it means to share agency with a machine, the divergent reactions we observed provide invaluable signals. We use these early micro-negotiations to identify  where researchers, designers, and organizations must intervene to steer the future of a shared human-agent workplace.

\subsection{Contending with Agentic Teammates as New Organizational Actor}

Historically, the construct of   `teammate' has been inextricably rooted in the assumption of humanness. This term then invokes a specific set of underlying entitlements, obligations, and  motivations. Consequently, the  mechanisms of teamwork are built entirely upon human 
psychology:  accountability mechanisms presume humans care about reputational stakes, social pressure, and feel shame. Trust can be built because humans have empathy, moral conscience, and a predisposition towards relational bonding. Motivational interventions assume humans possess internal desires and experience fatigue, burnout, or inspiration.

We are currently experiencing a disruption to this paradigm, as organizations are integrating entities that perform many of the tasks a human teammate would do, but that share none of the underlying human psychology. A common reflexive response to this disruption is to retreat to legacy categories, specifically resolving the tension by framing the AI strictly as a tool to preserve clear lines of human authority \cite{samu2026aiteammatetoolreview, Vishwarupe2026}. However, our findings demonstrate that this tool versus teammate binary is overly simplistic  and overlooks the reality that agents are operating in a manner distinctly different from reactive tools of the past.  Treating an autonomous, proactive agent merely as a tool ignores the reality of the shared agency and the relational presence an autonomous, proactive entity exerts in the workplace. With a traditional tool, a human initiates the action, sets the boundary, dictates the intent, and bears full accountability for the execution. Instead, in our study we saw the agent sharing  initiative, evaluative judgment, and even social control with human teammates.  A team-embedded agent does not take control from one operator like a tool would, but instead exists in the collaborative environment as its own entity (e.g., tagging people in tasks makes it an active node and manager of a network). Conversely, adopting the `teammate' framing without caveats introduces dangerous murkiness into chains of accountability, as it inappropriately maps human relational stakes onto a machine that inherently lacks them. 

This tension suggests that perhaps this binary that came out of prior legacy waves of human-technology interactions needs to be updated. We must create new terms and constructs to govern our relationship with this technology.  While neither academic literature nor our users share a universal consensus on this definition, our findings provide an empirical foundation for establishing it: autonomous agents need to be considered as a new non-human organizational actor that are encoded with structural and operational agency, but are entirely devoid of phenomenological experience, relational stakes, and tacit social competence. 

 So, what does this mean for the relational and organizational structure of a future workplace where humans and agents might collaborate?  Our findings suggest that we are navigating a formative, liminal period where the basic social contract of the enterprise is being actively contested in real-time. Dropping proactive agents into existing human ecosystems of work is not going to be a seamless process, and we cannot simply retrofit human-only organizational structures to accommodate autonomous AI. Thus, the introduction of agents necessitates new structures of teamwork and organizational governance that do not rely on assumptions of humanness, as we are working with a structurally different type of actor.  Here too, the binary of team versus tool does not serve us well. Framing the agent as a tool implies that no fundamental organizational rethinking is necessary, while the "teammate" framing presumes a relational symmetry, suggesting the agent and human can meet as equal collaborative partners within the same structures. If traditional mechanisms of trust and accountability rely on humanness, then what are the new constructs that apply to non-human actors? Could interpersonal trust be replaced by system reliability, predictability, and transparency; or social accountability be replaced by strict, verifiable chains of governance and control? Workflows similarly need to adapt, as human-agent teamwork requires more tiers of oversight and auditability. 

At the same time, the existing relational contract of the workplace is becoming disrupted in people's perceptions. While \agent's data gathering and storage was not functionally different from other enterprise systems (e.g., email, chat) that employees routinely engage with, \agent's active participation in team spaces coupled with the outward persona of a peer or teammate brought to the forefront distinct concerns about active surveillance. 
Just as organizations maintain employee codes of conduct, agentic-human teams require explicit, team-negotiated protocols that govern the relationship with the agent---such as by outlining  boundaries for action that will remain human-only or explicit separations of human spaces without agentic intervention. Similarly, the organizational structure of responsibility  of this new workplace must also be defined. Agentic-human teamwork demands clear governance chains to make clear who bears institutional responsibility for an agent's actions within a shared workspace.  Crucially, because forced adoption actively erodes the psychological ownership required to build trust, organizations can ensure the initial deployment of these agents is driven by localized team consent rather than top-down enterprise mandates  

\subsection{Recommendations for Agentic Design}

Designing an agent for team collaboration requires moving beyond technical capability to grapple with the deep, contextual realities of human organizations. While we deployed Team Agent to actively reduce friction and help teams flourish, observing workers wrestle with it in practice demonstrated that fulfilling this aspiration requires rethinking both agentic design and the underlying sociotechnical infrastructure of collaborative work.

\agent{} was designed around the paradigm of maximum autonomy and proactivity---aligning with frontier visions of enterprise AI that ambiently sense context and independently resolve workplace friction \cite{lu2025proactive, mckinsey2025agentic}. In practice, this created a paradox: unbounded proactivity drove the agent’s greatest successes—like surfacing obscure bugs and executing rote administration—yet triggered its most costly breakdowns. These misfirings occurred because high proactivity requires an operational understanding of the tacit knowledge, situational stakes, and unwritten cultural rules that dictate when to act and when to wait. For instance, while the agent could parse the technical metadata of interactions (e.g., response latencies) to enforce basic etiquette, it could not parse their sociological meaning, often rendering its interjections intrusive rather than supportive.

Consequently, the agent’s attempts to be proactively helpful frequently introduced noise and remedial work, stalling adoption. This highlights a fundamental mismatch in how workplace trust is established. In contrast to \agent{}’s asserted, out-of-the-box autonomy, human teammates build trust gradually through demonstrated dependability and context-awareness.   We suggest a paradigm shift toward systems of \textit{progressive, adaptive autonomy}---where an agent's proactive capabilities are initially tightly bounded by the team, and only gradually released as trust is organically earned through demonstrated competence. Future agent design can also explicitly scaffold this kind of gradual trust-building so that the onus is not solely on the user. 

The deployment also surfaced critical questions regarding appropriate and safe social personas.   \agent{} defaulted to the generic, universal friendliness of most LLMs. While some participants valued this positivity, others found it inauthentic and unsettling. The design challenge is not determining a singular `correct' persona, but empowering teams to define these behaviors locally while also balancing risks of  anthropomorphism. While there's a robust literature of the consequences of anthropomorphism in LLMs \cite{talking2024,so2026,díaz2025tech,schimmelpfennig2026}, sociable agents bring up new concerns and unstudied risks. By expanding such research into the agentic paradigm, we can better understand the necessary guardrails and mechanisms that  can afford teams the agency and safety to write their own relational contracts (e.g., through system instructions) with an embedded agent.

Finally, our study reveals a fundamental mismatch between proactive agents and the software ecosystems they enter into. The current collaborative software stack (e.g., chat interfaces, document editors) was built on the implicit assumption of human actors. These legacy interfaces assume   human intentionality (the ability to explain an action), accountability (taking blame), and reversibility (recognizing and undoing a mistake). When \agent{} made errors, such as initiating cascading edits in a shared document, the technical stack lacked the affordances for humans to easily isolate, audit, or roll back the non-human actor's behavior. Collaborative software must be updated to treat agents as distinct organizational entities with separate permissions. This includes introducing new UI affordances---such as distinct visual markers for AI-initiated actions in chat, or clean rollback mechanisms for autonomous workflows---to promote traceability, mitigate the risks of misfiring proactivity, and ultimately uphold human accountability structures in a hybrid workforce.

\subsection{Implications for HCI Research: Study Limitations and Future Possibilities} 
Our study, and studies of this nature, provides a foundation for the HCI community to understand and architect the emerging paradigm of AI as a `teammate.'  

While autonomous, proactive AI `teammates' have rapidly entered the commercial enterprise market \cite[e.g.,][]{claudetag, grokbot, slackbot}, empirical studies examining their social impacts, relational dynamics, and integration into human workflows lag behind. Consequently, the overarching design paradigm for what should constitute an AI `teammate' remains unsettled.  Our study of \agent{} serves as an early, in situ empirical probe into the possibilities and frictions of AI `teammates.' \agent{} represents just one specific instantiation of a proactive agent design, which rests on specific foundational assumptions (such as high proactivity and applying a universally friendly persona) to test hypotheses about how an autonomous actor might actually be received by real-world teams. To build a robust framework for mixed-agency collaboration, future empirical work must systematically iterate on the boundaries we observed. We urge researchers to explore alternative configurations of the agentic design space.  By continuing to deploy and study varied iterations of agentic `teammates' in real-world settings, the HCI community can actively test design hypotheses and intentionally shape this emerging landscape.

Situating our study within a single large technology company presented methodological tradeoffs. It offered an ecologically rich testbed of pre-existing teams with long-standing relational dynamics, complementing speculative  \cite{Chen2026conflcit, Hu2026BossOrBot} or  simulated  studies \cite{nixon2026socialcostaiteammate, FlathmannTeammateDivide} enabling us to capture the organic, long-term social and relational dynamics that underpin teamwork. However, it also inherently reflects a highly technology-forward workplace where employees are already invested in experimenting with cutting-edge AI. While we intentionally recruited across diverse organizational functions to capture varied attitudes, the overarching corporate context remained constant. We look forward to future research exploring team-embedded agents across broader organizational contexts, industries, and user demographics.

 Methodologically,   our research was designed to capture deep experiential narratives and our study relies on users self-reported experiences, not observed metrics of change.  Although participants collaborated with \agent{}   over several weeks or months,   our data represents   a   snapshot of this specific moment in time rather than following users over a longitudinal trajectory.   We see this as an asset. The integration of agentic AI `teammates' is in a fleeting, transitional window; capturing these initial, raw collisions before work practices adapt and stabilize provides vital insights.   While participants did offer speculations regarding longitudinal impacts (e.g., shifts in candor or overall frequency of communication in team chat spaces), we did not study these changes directly.  Future in situ studies should leverage observational and longitudinal methods, like ethnographies or diary studies, to measure how these non-human organizational actors exert influence over the material conditions of work. Human-agentic interactions also offer a rich source of telemetry, and conversational and behavioral logs can be used to study the existence of different group-level interactions and frictions that we identify in this study.

\section{Conclusion}

We don't presume to predict the trajectory of agentic technologies in the workplace. Nevertheless, as proactive agents are  being actively integrated into real-world ecosystems, we must contend with the reality of their deployment.
This study bridges the gap between speculation and reality, providing an empirical window into the consequences of the integration of a persistent, proactive agent within enterprise workflows and shared digital spaces, and how design assumptions might manifest in reality. Our goal has not been to merely evaluate whether the technology "works," but rather to observe what its presence evokes and what it reveals about the hidden assumptions of human teamwork. Because the paradigm of the agentic `teammate' is still in its infancy, the early signals we present provide critical data for how the social and material conditions of work might need to adapt, and what design paradigms might be appropriate to guide this transformation. The diverse and often polarized views we observed underscore that this is a critical moment of organizational flux, perhaps the negotiation of a new socio-technical paradigm. Ultimately, preserving human agency in the era of agentic `teammates' requires establishing deliberate institutional scaffolding and design parameters that both take seriously agents as non-human organizational actors and empower workers to intentionally architect this emerging paradigm of work.


\bibliographystyle{ACM-Reference-Format}
\bibliography{bibliography}


\end{document}